\documentclass[aps,prl,twocolumn,a4paper,10pt,notitlepage,footinbib,superscriptaddress,showpacs]{revtex4-1}%
\usepackage[english]{babel}
\usepackage[T1]{fontenc}
\usepackage{amssymb,amsmath,amsfonts}
\usepackage{textcomp}
\usepackage{graphicx,color}
\usepackage[dvipsnames]{xcolor}
\usepackage[utf8x]{inputenc}
\usepackage{float}

\usepackage{tabularx}

\newcommand{\oneD}{\rm{1D}}%
\newcommand{\threeD}{\rm{3D}}%

\begin{document}

\title{A nonequilibrium strategy for fast target search on the genome}
\vspace{-0.2 cm}
\author{F. Cagnetta}
\affiliation{SUPA, School of Physics and Astronomy, University of Edinburgh, Edinburgh EH9 3FD, United Kingdom}
\author{D. Michieletto}
\affiliation{SUPA, School of Physics and Astronomy, University of Edinburgh, Edinburgh EH9 3FD, United Kingdom}
\affiliation{MRC Human Genetics Unit, Institute of Genetics and Molecular Medicine, University of Edinburgh, Edinburgh EH4 2XU, UK}
\affiliation{Department of Mathematical Sciences, University of Bath, North Rd, Bath BA2 7AY, United Kingdom}
\author{D. Marenduzzo}
\affiliation{SUPA, School of Physics and Astronomy, University of Edinburgh, Edinburgh EH9 3FD, United Kingdom}

\begin{abstract}
\vspace{-0.3 cm}
Vital biological processes such as genome repair require fast and efficient binding of selected proteins to specific target sites on DNA. Here we propose an active target search mechanism based on ``chromophoresis'', the dynamics of DNA-binding proteins up or down gradients in the density of epigenetic marks, or colours (biochemical tags on the genome). We focus on a set of proteins that deposit marks from which they are repelled -- a case which is only encountered away from thermodynamic equilibrium.  For suitable ranges of kinetic parameter values, chromophoretic proteins can perform undirectional motion and are optimally redistributed along the genome. Importantly, they can also locally unravel a region of the genome which is collapsed due to self-interactions and ``dive'' deep into its core, for a striking enhancement of the efficiency of target search on such an inaccessible substrate. We discuss the potential relevance of chromophoresis for DNA repair.
\end{abstract}

\maketitle

Within the crowded nucleus of eukaryotic cells, it is vital that selected proteins and enzymes can locate their target on chromatin---the complex of DNA wrapped around histone octamers~\cite{Cortini2015}---within minutes~\cite{Rudolph2018,Chen2018} of a specific stimulus. DNA lesions, for instance, occur several thousands of times a day in every cell~\cite{Smeenk2013,Jackson2009}: the requirement for speed of the relevant repair machinery is thus not negotiable.

Passive mechanisms are unlikely to offer the required efficiency: a protein exploring a human chromosome---average size $\simeq 10^8$ base pairs (bp)---via 1D diffusion along the DNA ($D_{\oneD}< 10 \rm{kbp}^2\rm{s}^{-1}$) would take over $10$ years to find a single target. Purely 3D diffusion within the human nucleus, whose typical size is $\sim 10 \rm{\mu m}$, is equally impractical. Its limits are apparent by using Smoluchowski's prediction for diffusion-limited reaction rates~\cite{Richter1974}, $k_{\threeD} \simeq 4\pi D_{\threeD} a \simeq 10^7 \, \rm{M}^{-1} \rm{s}^{-1}$, calculated with $D_{\threeD}=10^{6} \rm{nm}^2\rm{s}^{-1}$  and $a \sim 10\rm{nm}$, as relevant \emph{in vivo}~\cite{Baum2014,Caragine2018}. This estimate leads to sufficiently short searching times ($1\text{\textendash}10\rm{s}$) only for high concentrations of searching proteins---$10^5\text{\textendash}10^6$ per cell, or $10^{-1}\text{\textendash}1\rm{\mu M}$.

Some of the components of the repair machinery are indeed highly abundant~\cite{kraus2003parp}: it is unclear, however, how they can access the collapsed chromatin conformations which are typically observed in the nucleus~\cite{mirny2011fractal,Boettiger2016,Michieletto2017nar}.
The combination of alternate rounds of 3D diffusion in the nucleus and 1D diffusive sliding on the DNA, or \emph{facilitated diffusion}~\cite{Berg1981,Kolomeisky2011,Slutsky2005,sheinman2012}, can also lead to faster search, but the speedup is at most one order of magnitude~\cite{Halford2004,Brackley2012,Brackley2013b} and requires a fully accessible substrate. These considerations lead us to conjecture that energy-consuming processes may be involved in the location of DNA lesions, as well as in other functional enzyme-DNA interactions aimed at the quick exploration of either swollen or collapsed chromatin conformations.

\begin{figure}[t!]
	\centering
	\includegraphics[angle=-90,width=0.45\textwidth]{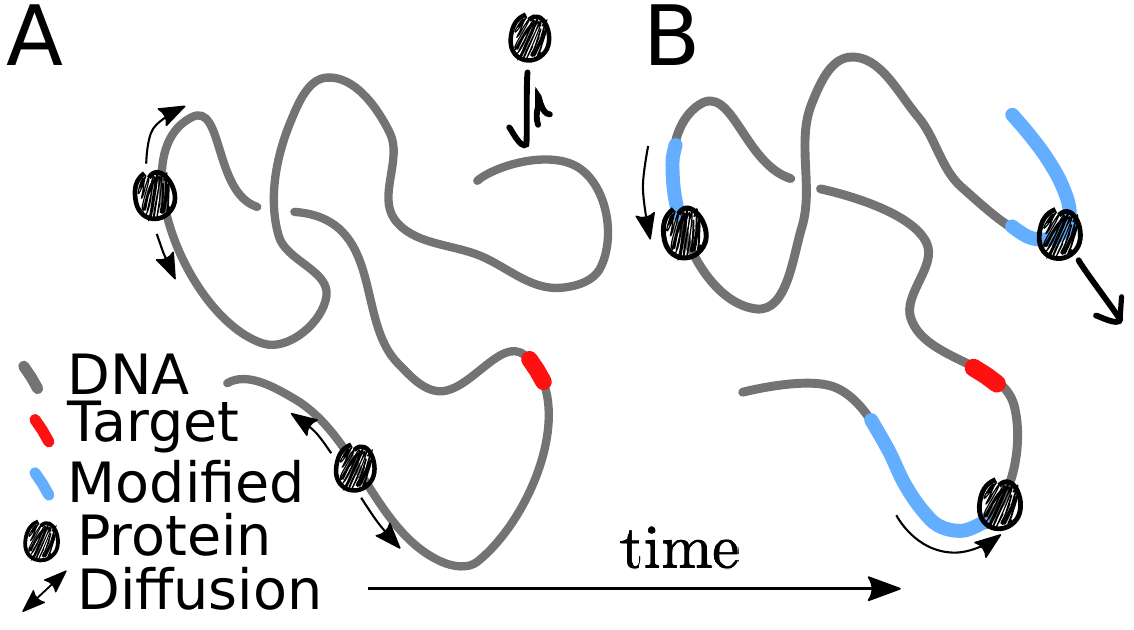}
	\caption{\textbf{Chromophoretic search.} \textbf{A} Proteins bind and diffuse along a fluctuating chromatin substrate (grey) searching for a target (red). \textbf{B} Proteins deposit epigenetic marks (cyan) along the substrate at rate $k_{\rm{on}}$. The repulsion between proteins and the deposited marks results in directed motion and nontrivial collective behaviour.}
	\vspace{-0.5cm}
	\label{fig:Model3D}
\end{figure}

In this Letter we introduce the concept of \emph{chromophoresis}---the spontaneous motion of DNA-binding proteins as a result of a self-produced pattern of chemical \emph{marks}---modifications, such as acetylation or methylation, of histone proteins that, together with DNA, form chromatin. The prefix chromo- is chosen because, in our model, DNA-binding proteins move along colour gradients (as marked histones can be thought as having a different colour than unmarked ones, Fig.~\ref{fig:Model3D}), and also suggests that phoresis occurs along chromatin.
As the marks deposited provide a layer of inheritable information beyond the DNA sequence, they are referred to as \emph{epigenetic}. In the context of epigenetics, biophysical models normally consider a positive feedback loop between the released epigenetic marks and the protein dynamics, leading to accumulation and pattern formation~\cite{Dodd2007a,Michieletto2016prx,Erdel2016,Jost2014a}. However, assuming an energy input allows also for \emph{negative} feedbacks, whereby a protein deposits a mark from which it flees.

An exemplary instance of negative feedback, which is also involved in DNA repair, is PARylation: the addition of Poly ADP-ribose (PAR) chains on histone proteins~\cite{Smeenk2013,kraus2003parp}. PARylation is know to decrease the local affinity to chromatin-binding protein; and to facilitate the recruitment of reparing enzymes at the lesion~\cite{Gibson2012,chaudhuri2017multifaceted}. Its role in the location of the lesions, instead, is still under debate.
In general, the scenario we propose is reminiscent of chemorepulsion in active matter, where it leads to coordinated motion~\cite{,Saha2014,liebchen2015,cagnetta2018}. As we shall show, negative chromophoresis also yields nontrivial patterns and it provides a generic nonequilibrium mechanism for fast target search on chromatin. Intriguingly, the mechanism works even on a collapsed globule, where the target may not be immediately accessible to diffusive searching proteins.

\paragraph{Model --}
We consider a 3D model for protein chromophoresis on a chromatin fibre. The latter is built, following a well-established description of eukaryotic chromosomes {\it in vivo}~\cite{Rosa2008,Brackley2013,Jost2014B,Barbieri2013,Michieletto2016prx,Giorgetti2014}, as a flexible bead-and-spring polymer of length $M$. Each bead represents a set of nucleosomes, and we set the bead size $\sigma$ to $1-3$ kbp, or $10-30$ nm. Chromophoretic proteins are represented as $N$ spherical beads with viscous friction coefficient $\gamma$ and are assumed, for simplicity, to have also size $\sigma$. The system is immersed in a heat bath at temperature $T$, and the equations of motion are solved by using LAMMPS in Brownian Dynamics mode~\cite{Plimpton1995} (see SI). In the absence of any chromophoretic process, proteins bind to the fibre non-specifically with affinity $\epsilon$, modelled as a truncated Lennard-Jones (LJ) potential. Provided $\epsilon$ is comparable to $k_BT$, proteins can slide on chromatin, with an effective diffusion coefficient $D_{\rm 1D}$ (Fig.~\ref{fig:Model3D}A, see also Ref.~\cite{Brackley2012}). Larger $\epsilon$ instead leads to cluster formation via the \emph{bridging-induced attraction}~\cite{Brackley2013}. 

The proteins we consider here deposit an epigenetic mark on the fibre bead they are bound to at a rate $k_{\rm on}$ (Fig.~\ref{fig:Model3D}, B). This mark, in turn, abrogates the attraction of the protein to the marked beads. Marks are spontaneously lost at rate $k_{\rm{off}}$, modelling random or active removal. This model harbours a negative feedback, as the mark released by the proteins raises, rather than lower, the potential energy describing fibre-protein interaction.
Hence, unlike the case of positive feedback~\cite{Michieletto2016prx,Michieletto2017nar}, this system cannot be described by an effective equilibrium model.
To understand the dynamics that can originate from these microscopic rules, we first consider a simpler 1D model that neglects spatial structure and fluctuations of the chromatin fibre.

\begin{figure}[t!]
	\centering
	\includegraphics[angle=-90,width=0.45\textwidth]{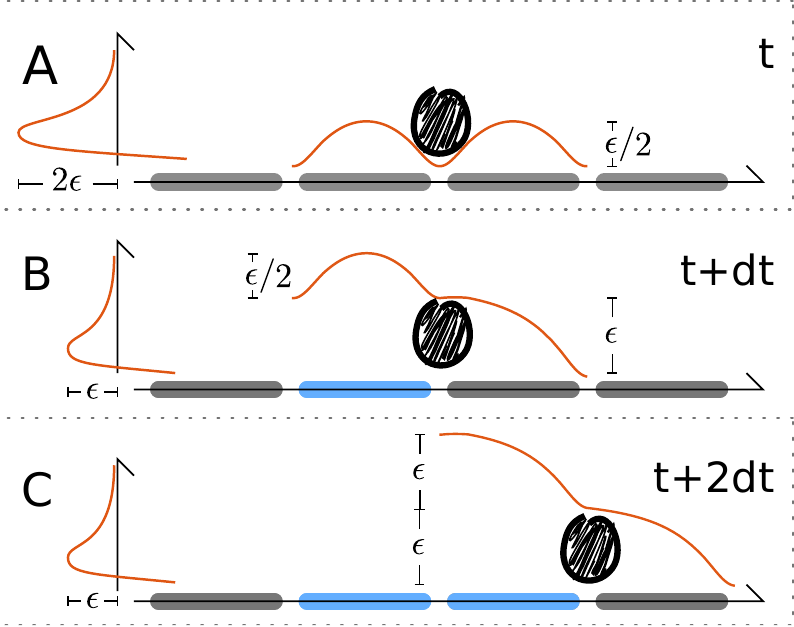}
	\caption{\textbf{Simplified 1D model.} \textbf{A} Potential landscape seen by a protein binding to an unmarked (grey beads) region of the chromatin fibre. \textbf{B} Potential after deposition of a mark (blue bead). \textbf{C} Potential after the protein hops one bead to the right and deposits another mark.}
	\label{fig:Model1D}
\end{figure}

\paragraph{1D approximation --}
As a first approximation, the chromatin fibre can be treated as a straight line.
The potential landscape generated by the LJ interaction with the fibre determines the protein dynamics, and is substantially easier to compute for a 1D substrate (see Fig.~\ref{fig:Model1D} and SI). In the absence of any mark, a protein sits between two adjacent fibre beads so as to minimise the potential energy (Fig.~\ref{fig:Model1D}A). The protein can escape the well in the direction orthogonal to the fibre (vertical axis in Fig.~\ref{fig:Model1D}). The escape rate $r_{\rm esc}$ can be computed, for $\epsilon\,{\gtrsim}\,k_B T$, as a Kramers problem~\cite{kramers1940}, and scales as $\sim e^{-2\epsilon/(k_BT)}$.  Additionally, a barrier $\epsilon/2$ obstructs the protein motion parallel to the fibre (horizontal axis in Fig.~\ref{fig:Model1D}), i.e. the thermal diffusion between adjacent potential wells. 
Kramers' theory (see SI) yields the effective hopping rate for the symmetric random walk the protein performs on the fiber as, $q\,{=}\,\frac{A}{4}\frac{\epsilon}{1+e^{\epsilon/2k_BT}}$, with $A\,{\simeq}\,10.6/2\pi$ a numerical factor depending on the potential curvature. Note that, unless otherwise stated, we set $k_BT\,{=}\,\gamma\,{=}\,\sigma=1$ in what follows, so that dimensional formulas for rates can be obtained by multiplying those we give here by $k_B T/(\sigma^2 \gamma)$ (see SI).

The diffusion coefficient of the 1D diffusive sliding is then $D_{1D}\,{=}\,q$. In our model, however, a bound protein deposits an epigenetic mark on one of the neighbouring chromatin beads which, by silencing the LJ attraction, reshapes the potential landscape (Fig.~\ref{fig:Model1D}B) and drives the model away from equilibrium.
While the barrier over the marked bead remains unaltered, the one over the unmarked one is tilted, becoming a declivity of size $\epsilon$. From Kramer's theory, the rate at which the protein slides down the declivity is $q_{+}=B\epsilon$, with $B\simeq 22.5/ 2\pi$. As $q_+/q \simeq 8\left(1+e^{\epsilon/2k_BT}\right)$, we expect the protein to slide downhill with near-one probability (recall $\epsilon\gtrsim k_BT$ for Kramers' theory to hold). 

If the mark-deposition rate $k_{\rm on}$ is large enough, the protein is likely to mark the underlying bead, thus ending up in the configuration depicted in Figure~\ref{fig:Model1D}C. As there are two marked beads upstream of the protein, the barrier over the marked bead changes: the rate of sliding downhill remains $q_+$ while that of hopping backwards changes to $q_-= C \epsilon e^{-\epsilon/k_BT}$, with $C\simeq 7.5/2\pi \gtrsim 1$. 
Therefore, in a typical microscopic sequence of events, a chromophoretic protein binds to the substrate, then it randomly marks one of the two beads on either side and becomes attracted to the other. In doing so, it enters a \emph{running} state, whereby it slides forward towards the unmarked portion of the fibre at rate $q_+ \sim \epsilon$ or hops backward onto the marked segment at rate $q_-\sim \epsilon e^{-\epsilon/k_BT}$. A backward hop would end the running state, forcing the protein off the fibre, hence renormalising the escape rate to $r'_{esc} = r_{esc} + q_- \simeq 2q_-$. The relevant lengthscale of this process is the ``run length'', i.e. the chromatin segment that the protein explores before detaching. This is given by $l_{\rm run}\sim v/r'_{\rm esc}$, where $v \propto q^+$. More precisely,
$ l_{\rm run} = B/2C e^{\epsilon/k_BT} \simeq 3/2 e^{\epsilon/k_BT}$.

Mark evaporation does not change the picture, unless occuring at rate $k_{\rm off} > q_+$, which we never consider here~\footnote{As $k_BT=1$, we need $\epsilon>1$ for Kramer's picture to apply, hence $q_+ \gtrsim 3$.}.
It is required, instead, that $k_{\rm on} \gg q_+$, though 3D simulations show the running state exists down to $k_{\rm on} \sim q$. Reinstating dimensional factors, this translates into $k_{\rm on} \,{>}\, D_{\oneD}/\sigma^2$, or $k_{\rm on} \,{>}\,\rm{s}^{-1}$ for $D_{\oneD}\sim 10^{-3}$ $\mu$m$^2$/s, a bead size $\sigma=30$ nm and $\epsilon \,{\gtrsim}\, k_BT$, as for proteins on chromatin. This rate of post-translation modification is compatible, albeit slightly faster, than that of typical modifications: for instance, $k_{\rm on} \simeq\rm{min}^{-1}\,{-}\,\rm{s}^{-1}$  for acetylation or phosphorylation~\cite{Sun2003}.

\begin{figure}[t!]
	\centering
	\includegraphics[angle=-90,width=0.45\textwidth]{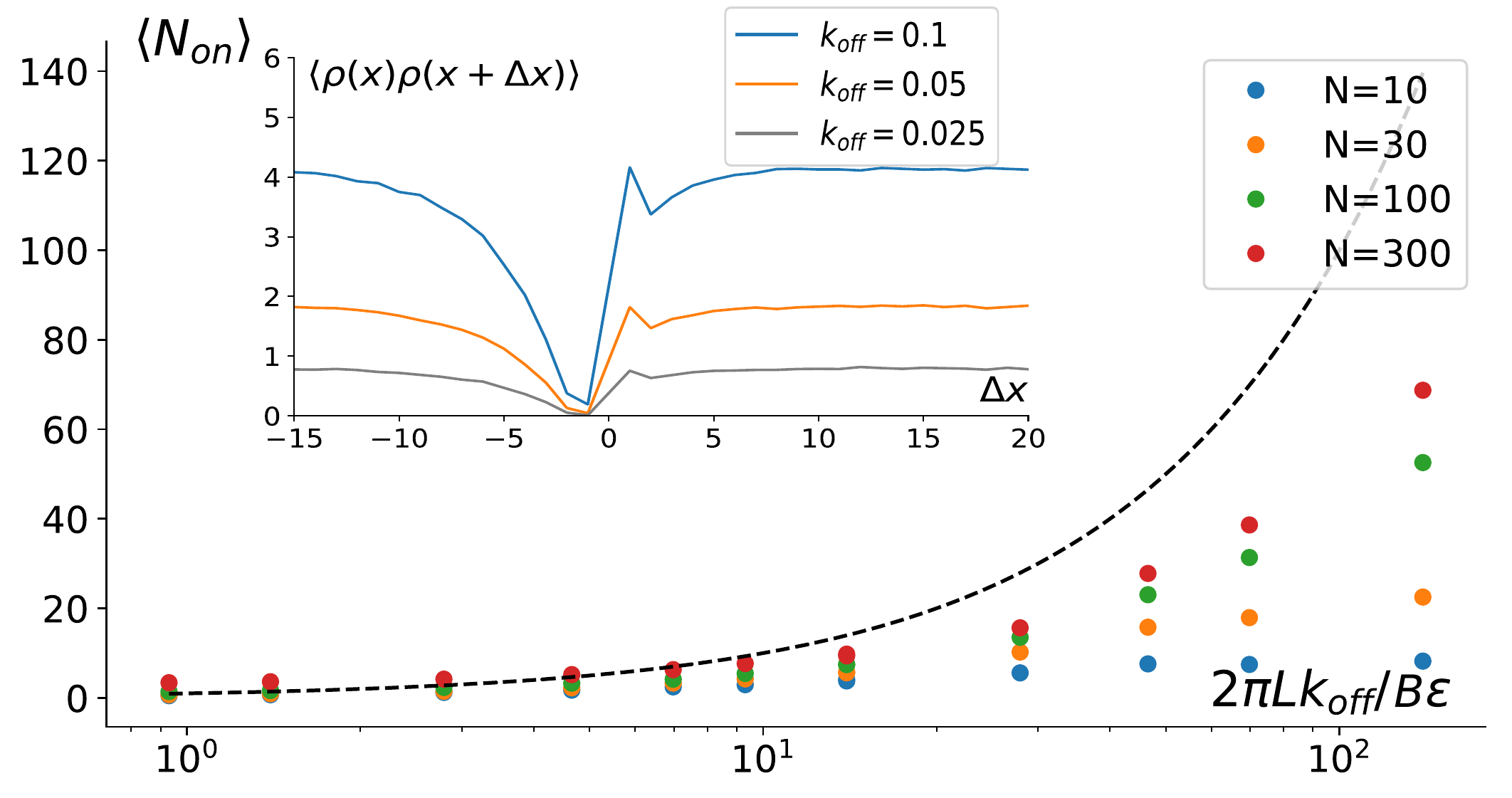}
	\caption{\textbf{Chromophoretic collective behaviours.} Average number of chromophoretic proteins bound and moving along the substrate for different values of total protein copy number $N$ and mark removal rate $k_{\rm off}$. The black dashed line marks the limiting protein number $L/l_{trail}$ discussed in the text: it provides an upper bound for $\left\langle N_{on}\right\rangle$. The inset shows the two-point correlation function in the direction of the protein motion (to the right in the figure).}
	\label{fig:Stats1D}
\end{figure}

The unidirectional motion of a single protein accelerates target search substantially, by enlarging the distance covered while bound to the substrate. 
In addition, multiple chromophoretic proteins bound on the same fibre interact with each other via the trails of epigenetic marks left on the substrate. 
This effect is manifest in the pair correlation function, i.e. the average density profile seen by a running protein (Fig.~\ref{fig:Stats1D}, inset). The downstream peak at short distance is due to collisions with proteins moving in the opposite direction, and the upstream dip to the epigenetic trail and consequent protein depletion. Due to this forward-backward asymmetry, the resulting effective interaction breaks the action-reaction principle (e.g., a particle in the wake of another is repelled by the latter without affecting its motion), underscoring the nonequilibrium nature of the model.

The inter-particle interactions are thus controlled by the epigenetic mark dynamics: this provides an avenue to set up a cooperative search strategy, which is unavailable to conventional facilitated diffusion. Due to the trail-mediated exclusion between proteins, the average number of proteins bound at any time does not exceed $L/l_{trail}$, where $l_{trail} \sim v/{k_{\rm off}}$ is the average trail length. We therefore expect chromophoresis to suppress stochastic fluctuations in the relative distance between neighbours (see SI), leading to \emph{hyperuniform} spreading along the substrate~\cite{Torquato2018hyperuniform}. A direct consequence of this is a faster search, as each protein only needs to scan a range $\sim l_{trail}$ and is unlikely to bind to a segment which has already been scanned. Simulations confirm that the average number of bound proteins is controlled by $k_{\rm off}/\epsilon$ (Fig.~\ref{fig:Stats1D}). Biologically, $k_{\rm off}$ can be modulated in response to endogenous or external stimuli for many epigenetic marks~\cite{Sun2003}. 

\begin{figure}[t!]
	\centering
	\includegraphics[angle=-90,width=0.45\textwidth]{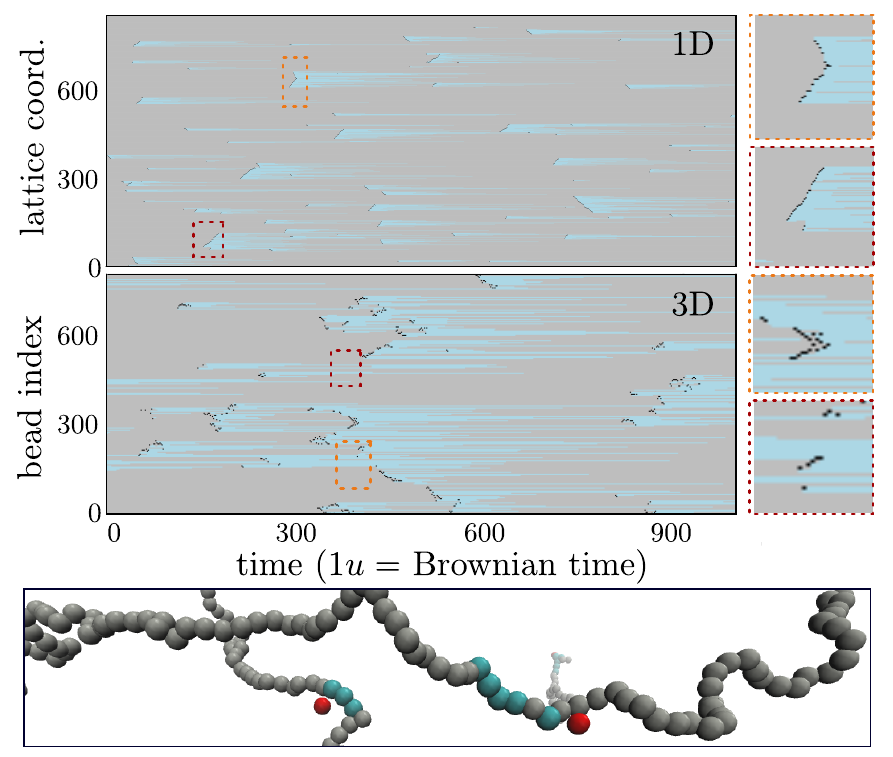}
	\caption{\textbf{Kymographs of chromophoretic proteins}. Kymographs showing the epigenetic mark dynamics. \textbf{A} 1D model with $M=1000$, $\epsilon=2$, $k_{\rm off}=0.01$, $k_{\rm on}=10$ and $N=20$ proteins which, when not on the fiber, re-bind to it at rate $0.1$. \textbf{B} 3D model with with $M=1000$, $\epsilon=4$, $k_{\rm off}=0.01$, $k_{\rm on}=1$ and $N=20$ proteins in a $L=50$ cubic box. \textbf{C} Snapshot from 3D simulations showing chromophoretic proteins (red) and epigenetic marks (cyan) on chromatin (grey).}
	\label{fig:Kymographs}
	\vspace{-0.5 cm}
\end{figure}

\paragraph{3D Model --} 
We now discuss the case where the chromophoretic dynamics occurs on a 3D fluctuating chromatin fibre. We first focus on parameters for which the fibre is swollen (Fig.~\ref{fig:Kymographs}, Suppl. Movie 1). The proteins dynamic can be quantified via kymographs, showing the local epigenetic state of the polymer (grey=unmarked or cyan=marked) overlaid with protein positions (black) versus time. Both 1D and 3D systems display the same dynamical features, such as collisions and trail-mediated dissociations (Fig.~\ref{fig:Kymographs}A,B).  


The eukaryotic genome \emph{in vivo}, however, is not a swollen fibre but is understood as a confined and microphase separated polymer~\cite{lieberman2009comprehensive,boettiger2016super,Brackley2016NAR,Michieletto2016prx,Brackley2017biophysj,Michieletto2017nar,erdel2018formation,chiariello2016polymer,jost2014modeling}. In particular, a locally collapsed state is a typical representation of a \emph{heterocromatic}, or transcriptionally silent, genomic region~\cite{Boettiger2016,Michieletto2016prx}. Whilst standard facilitated diffusion studies are normally carried out in swollen conditions~\cite{Mirny2009a,Brackley2012}, lesions and single- or double-stranded DNA breaks may often be buried within collapsed and inaccessible heterochromatic regions

Thus, to explore a regime of target search relevant for DNA lesion repair \emph{in vivo}, we perform simulations with a number of protein bridges that fold the polymer substrate into a collapsed globule~\cite{Brackley2013pnas,Barbieri2012,LeTreut2016}---modelling, for instance, multivalent HP1 proteins associated with heterochromatin~\cite{Kilic2015}. Once the chromatin fibre is folded by these abundant bridges, we release a trace amount of chromophoretic searchers. The two species of proteins interact sterically and, for simplicity, each has the same binding affinity for unmarked chromatin. We also assume that neither bind to the epigenetic mark deposited by the chromophoretic species. Inspection of the simulations shows that, strikingly, chromophoretic searchers can disrupt bridging-induced collapse and locally open the chromatin fibre. It is notable that similar phenomenology is observed during DNA repair as large chromatin regions surrounding DNA breaks swell~\cite{Seeber2018}. 

\begin{figure}[t!]
	\centering
	\includegraphics[width=0.45\textwidth]{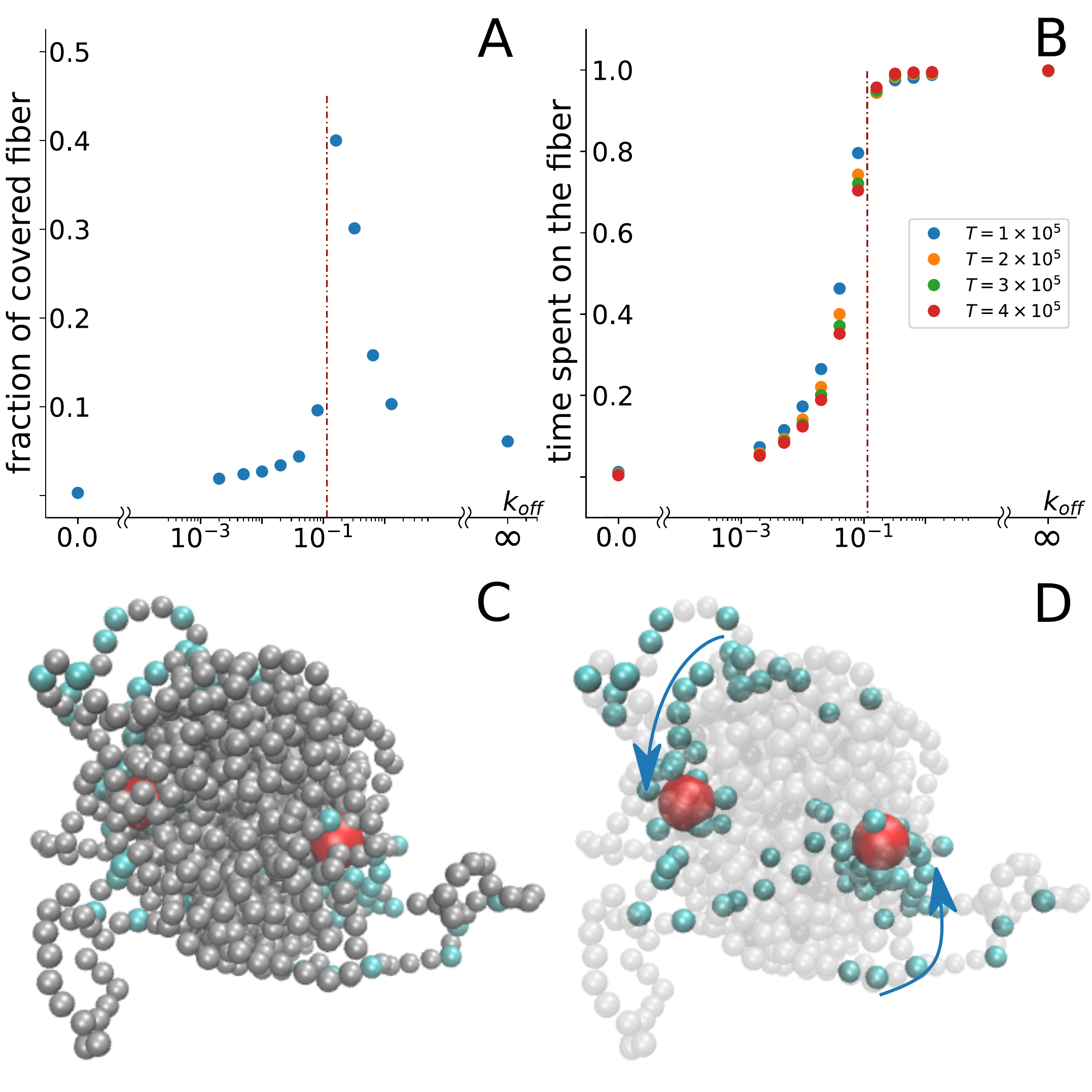}
   \vspace{-0.5 cm}
	\caption{{\textbf{Chromophoretic search on a collapsed substrate.}} \textbf{A} Average fraction of the fibre visited in a single binding-unbinding, or ``diving'', event as a function of $k_{\rm off}$ for $k_{\rm on}=0.1$ and $\epsilon/k_BT=5$. 
Results are averaged over several diving events and $10-20$ independent simulations.
\textbf{B} Fraction of time spent on the fibre over the total simulation time $\tau/T$ for different observation times $T$.
In \textbf{A} and \textbf{B} the red dot-dashed line highlights a critical $k_{\rm off}$ marking the value at which the fraction of covered fibre is maximal and there is a transition in the behaviour of residence time $\tau/T$.
	\textbf{C,D} Snapshots of two chromophoretic proteins (red) ``diving'' into a globule while pushed by their own trail (cyan).}
	\vspace{-0.3 cm}
	\label{fig:Divers}
\end{figure}

In order to quantify the efficiency of chromophoretic search, we monitor the fraction of beads that are visited each time a chromophoretic protein binds the substrate. Remarkably, we discover that there is a non-monotonic behaviour as a function of $k_{\rm off}$ (Fig.~\ref{fig:Divers}), which can be explained as follows. For $k_{\rm off} \rightarrow \infty$ the epigenetic marks evaporate immediately and the searchers only stick to the surface of the polymer globule: this limit is analogous to conventional facilitated diffusion, where a buried target would remain inaccessible to the searchers. In the opposite regime, $k_{\rm off} \rightarrow 0$,  the fibre swells but the searchers fail to remain attached for long because of the large fraction of non-sticky epigenetic marks. In both these limits, therefore, the fraction of beads visited for each binding event tends to $0$ (panel A of Fig.~\ref{fig:Divers}).

In the regime of intermediate $k_{\rm off}$ 
we instead observe a qualitatively different phenomenon:
searchers can be seen ``diving'' into the globule during simulations (Fig.~\ref{fig:Divers} and Suppl. Movie 2), by creating a local opening made of marked beads that slowly turn to unmarked. During the turnover time, the searcher is likely to be driven further inside the globule (i.e., to dive) as, on average, the protein sees a gradient of unmarked beads towards the interior. This gradient is actively maintained by the deposition of epigenetic marks, and fuels the descent of the chromophoretic searchers into the core of the globule. 

Once a searcher has dived deep enough into a globule it may remain trapped for a long time due to the large density of unmarked beads which it can stick to. During this time it can explore a large fraction of the polymer contour length by constantly churning the inside of the polymer globule. As a result, the optimum turnover rate $k_c$ marks a cusp in the fraction of fibre visitided per dive as a function of $k_{\rm off}$.
We further find that for $k_{\rm off}\,{>}\,k_{c}$ searchers spend a very long time attached to the fibre, but cannot make much progress inside the core due to steric effects, whereas for $k_{\rm off}\,{<}\, k_{c}$ the time spent on the fibre after binding is finite (i.e., it tends to zero for sufficiently long simulations, Fig.~\ref{fig:Divers}B). 

\paragraph{Conclusions --} In summary, we have proposed a novel nonequilibrium mechanism for target search within the genome. Inspired by the deposition of epigenetic marks on chromatin and consequent response of certain proteins to the gradient of such marks, we dub this mechanism ``chromophoresis''. In this work we focussed on a {\it negative} feedback between marks and proteins, so that the proteins are repelled from the mark they deposit. We discover that, if mark deposition is sufficiently fast, a single chromophoretic protein can perform unidirectional motion on chromatin, while multiple proteins interact via epigenetically-mediated repulsion, as a result of which they spread out along the fibre with suppressed 1D density fluctuations. Thus, we found chromophoresis to provide a generic pathway for accelerated target search, especially in cases where the chromatin fibre collapses into a globular configuration, as in a large fraction of the human genome. Under this condition, we proved the existence of an optimal evaporation rate of epigenetic marks for which the exploration of the fibre is fastest. Close to the optimal condition the proteins can locally untangle the globular chromatin and dive into its core, which is inaccessible to simple passive searchers performing facilitated diffusion.

In addition to the intriguing physics, chromophoresis is potentially relevant in the context of chromatin PARylation. Proteins of the PARP family, which are the chromatin-binding proteins responsible for PARylation, are recognised as a key part of the repair machinery: as such, they need to locate DNA lesions~\cite{Smeenk2013}, which might be buried within collapsed chromatin globules. PARylation has been shown to swell chromatin \emph{in vitro}~\cite{Poirier1982} and is thought to affect the dynamics of PARP itself, as well as of other proteins, promoting their detachment from the fibre~\cite{Gibson2012}.
We therefore speculate that chromophoresis may provide a mechanism for PARP to locate DNA lesions, a process which is currently poorly understood---we hope our work can stimulate experiments to explore this possibility further.
Another experimental pathway that we are keen to explore is the bottom-up  design of synthetic chromophoretic proteins. 

FC acknowledges support from the Scottish Funding Council under a studentship. DMa and DMi thank the European Research Council (ERC CoG 648050 THREEDCELLPHYSICS) for funding. The authors thank M. R. Evans for valuable feedback on the manuscript.

\bibliographystyle{apsrev4-1}
\bibliography{library_dmi}

\end{document}